\documentclass[journal,twoside]{IEEEtran}
\usepackage{cite}
\usepackage{amsmath,amssymb,amsfonts}
\usepackage{algorithmic}
\usepackage{enumitem}
\usepackage{graphicx}
\usepackage{textcomp}

\title{MRNet: Multifaceted Resilient Networks for Medical Image-to-Image Translation}
\author{Hyojeong Lee, Youngwan Jo, Inpyo Hong, Sanghyun Park, \IEEEmembership{Member, IEEE}
\thanks{This research is supported by the National Research Foundation (NRF) funded by the Korean government (MSIT) (No. RS-2023-00229822).}
\thanks{Hyojeong Lee is with the Department of Artificial Intelligence, Yonsei University, 50 Yonsei-ro, Seodaemun-gu, Seoul, 03722, South Korea (e-mail: hyojoy@yonsei.ac.kr).}
\thanks{Youngwan Jo, Inpyo Hong, Sanghyun Park are with the Department of Computer Science, Yonsei University, 50 Yonsei-ro, Seodaemun-gu, Seoul, 03722, South Korea (e-mail: jyy1551@yonsei.ac.kr; hip9863@yonsei.ac.kr; sanghyun@yonsei.ac.kr).}
}

\begin{document}
\maketitle
\begin{abstract}
We propose a Multifaceted Resilient Network(MRNet), a novel architecture developed for medical image-to-image translation that outperforms state-of-the-art methods in MRI-to-CT and MRI-to-MRI conversion. 
MRNet leverages the Segment Anything Model (SAM) to exploit frequency-based features to build a powerful method for advanced medical image transformation. The architecture extracts comprehensive multiscale features from diverse datasets using a powerful SAM image encoder and performs resolution-aware feature fusion that consistently integrates U-Net encoder outputs with SAM-derived features. This fusion optimizes the traditional U-Net skip connection while leveraging transformer-based contextual analysis. The translation is complemented by an innovative dual-mask configuration incorporating dynamic attention patterns and a specialized loss function designed to address regional mapping mismatches, preserving both the gross anatomy and tissue details. Extensive validation studies have shown that MRNet outperforms state-of-the-art architectures, particularly in maintaining anatomical fidelity and minimizing translation artifacts. 
\end{abstract}


\section{Introduction}
\label{sec:introduction}
\IEEEPARstart{M}{edical} image-to-image translation aims to map relationships between different modalities. The main challenge is to transforming the distribution of a source image into that of a target image. As a result, generative adversarial network (GAN)-based pixel mapping has become widely used in the medical field for tasks such as image registration\cite{registrationexample}, CT noise denoising\cite{denoisingexample}, and MRI-CT/MRI translation. Advances in deep learning have made these applications feasible. Among the various tasks that deep learning supports, bidirectional conversion between modalities, such as MRI and CT is crucial due to challenges in acquiring single-modality due to the distinct image acquisition characteristics of medical images, such as the need to protect pregnant women from radiation or accommodate patients with implanted medical devices. Missing modalities may also occur due to clinical situations. and in such cases, modality conversion can be clinically beneficial.
In GAN-based image translation, the focus is on the generator's image generation process, with improvements to both the model and loss function for better results. A model designed to better capture the distribution of the source and target domains has been proposed\cite{changearchexample1}, along with a specialized loss function to enhance performance\cite{specialloss}.
However, in the medical domain, the goal is not just to generate an image resembling the target but to preserve the original underlying features during translation. It is important to recognize that images demonstrating stylistic similarity through latent vector representation may not always ideal for translation. Therefore, success translation requires preserving semantic content while identifying differences between source and target images . To address these challenges, we introduce the Multifaceted Resilient Network(MRNet), a novel medical image translation network. 

The novelties and contributions of this paper are summarized as follows : 
\begin{description}
    \item[$\bullet$] To the best of our knowledge, MRNet is the first model to integrate Segment Anything Model(SAM)-based frequency interpretation. The proposed framework leverages image encoding capabilities for medical image translation, focusing on preserving anatomical structures during MRI-to-CT conversion.
    \item [$\bullet$] Resolution specific feature fusion: We introduce a feature fusion mechanism that combines encoder features and SAM-encoded features across multiple resolutions, enabling the effective preservation of both local and global image characteristics.
    \item[$\bullet$] Dual-mask Framework: We introduce a multimask approach enhances feature selection during translation through dynamic mask generation and specialized loss terms. This method significantly improves the preservation of anatomical structures compared to conventional techniques. The proposed MRNet consistently outperforms state-of-the-art methods in MRI-to-CT and MRI-to-MRI image conversion.
\end{description}

\vfill 
\noindent\hrulefill
\vspace{5pt}
\noindent\makebox[\textwidth][l]{%
  \fbox{\parbox{\dimexpr 0.49\textwidth - 2\fboxsep - 2\fboxrule\relax}{%
      \footnotesize This work has been submitted to the IEEE for possible publication. Copyright may be transferred without notice, after which this version may no longer be accessible.%
  }}%
}
\vspace{-10pt}

\section{RELATED WORK}
\subsection{Image-to-Image Translation}
Image-to-image translation has evolved dramatically in recent years. The breakthrough came through conditional generative adversarial networks (cGANs) \cite{cgan}, which have demonstrated remarkable capabilities. cGANs improve existing GAN architectures by incorporating class-specific constraints. By directly passing the constraints as conditions, the generator can continuously challenge the discriminator to produce incredibly realistic cross-modality images. Pix2Pix \cite{pix2pix2017} took this concept further by directly conditioning the input images. If the latent vector-based generator's mapping function takes both a source image x and a random noise vector z to produce y, expressed mathematically as \( G: \{x,z\} \rightarrow \{y\} \). More recently, PPT \cite{ppt} introduced sophisticated patch-based contrastive learning techniques, achieving exceptional input-output consistency when used to analyze pathological image datasets.
When researchers cannot access paired domain data, they can define informative loss functions that minimize discrepancies between the target and generated images, leading to more precise results \cite{cut}. 

CycleGAN \cite{CycleGAN2017} proposed another architecture by enabling bidirectional image transformations with unpaired datasets.
This innovative approach to transforming style Y back to style X facilitates bidirectional image transformation using nonpaired datasets. To preserve the characteristics of the original image, cycle consistency loss was proposed during the conversion process.

CycleGAN-based approaches have been widely applied in medical image transformation tasks. However, cycle consistency loss makes it difficult to preserve diagnostic features, especially in medical applications. Although this method preserves global structures relatively well, it often struggles with fine anatomical details that are important for accurate diagnosis\cite{bad_cycleloss}. The most concerning aspect is that the generated artifacts may be mistaken for genuine pathological findings.

Additionally, recent studies on aspects of cycle consistency models have shown that they are adept at encoding hidden information via steganographic techniques \cite{cycleganstegomaster, Porav2019ReducingSI}. The use of this feature should be used cautiously in clinical settings where image integrity must be absolutely assured.

These limitations are important in medical imaging modalities where precise structural preservation of the image affects diagnostic accuracy. For example, tasks such as cancer detection and disease progression monitoring, where minor artifacts or distortions must be avoided, are involved. Given these considerations, it is best to use paired training data whenever possible in medical imaging applications.

Paired training offers distinct advantages. First, direct supervision allows for more accurate preservation of anatomical details and significantly reduces the risk of artifacts and structural distortions. In terms of validation, it also allows for comparison of the ground truth-to-ground transformation accuracy. Unpaired methods are still valuable when paired data are scarce or unavailable in medical imaging. However, developing high-quality paired datasets should be a priority to ensure optimal accuracy and reliability in clinical applications. Therefore, the focus should be on utilizing paired datasets whenever possible, while reserving unpaired methods for situations where paired data acquisition is impractical or impossible.

\subsection{SAM}
SAM(Segment Anything Model) \cite{sam2023} is a recently developed segmentation foundation model with strong generalization performance and applicability to diverse vision tasks. It comprises three main components: an image encoder, a prompt encoder, and a mask decoder.
The image encoder, built on the Vision Transformer (ViT) architecture, divides images into patches and encodes them to extract features. The prompt encoder processes various prompt types, such as points, boxes, and text, resolving ambiguities caused by insufficient prompt data. This innovative use of prompts has enabled the development of multimodal architectures. The mask decoder combines encoded image and prompt embeddings with output tokens to generate segmentation masks.

In this study, we utilized a pretrained base-size ViT SAM model trained on an 11M image dataset, enabling powerful feature selection through generalized representations. This approach reduces training time and enhances performance. We extract features using the ViT-based SAM image encoder and integrate them into the decoder branch.

\subsection{Unlocking Hybrid Synergy: Strengths of Vision Transformers and Convolutional Neural Networks}
Recent research has highlighted the distinct advantages of ViT in low-frequency signal processing, in contrast to convolutional neural networks (CNNs) that excel at high-frequency signal analysis \cite{vitgoodinlow1, vitgoodinlow2}. These differences suggest that developing a hybrid approach that combines the capabilities of ViT for SAM encoding with the analytical strengths of CNNs could ultimately improve overall processing efficiency.
To leverage this synergy effectively, we engineered a novel integration of SAM encoding within the decoder branches of the network, addressing the low-frequency processing limitations inherent to CNNs. Careful alignment of the feature layouts with the original network architecture was essential to ensure compatibility and maintain output integrity \cite{feature_same}. The encoder responsible for extracting the SAM-based features is built upon a stable pretrained model, facilitating coherent incorporation into the existing framework. By allowing residual connections across varying resolutions of the feature maps, we preserve the integrity of the input image structure.
This separate learning of the SAM based feature flow provides a comprehensive examination of the image and addresses the steganography challenges associated with GANs. In steganography, critical information is concealed within subtle patterns; however, this challenge can be managed effectively through noise introduction or blurring techniques \cite{wu2024stegogan}.

The importance of synthetic-based decoders in image transformation highlights the rationale for adopting hybrid models. The different frequency components of an image have a significant impact on its image characteristics. Low-frequency components are usually related to the overall style of the image, such as smooth texture and general appearance. High-frequency components are related to rapid changes in pixel values, resulting in details such as edges. \cite{frequency_texture}. 
This understanding is what shaped the design of the reconstruction decoder \(D_\theta\). By emphasizing clarity and structural integrity, it can reduce low-frequency components and reproduce important details more accurately. This strategic approach has improved the effectiveness of the image translation process. It leads to better results in converting esthetically pleasing and technically accurate images.
In addition to interpreting the frequency aspect, hybrid models that integrate the transformer with traditional CNNs have been applied in various fields. For example, there have been many efforts to improve local and global interactions.

Notable examples include TransUNet \cite{chen2021transunet}, which uses skip connections in CNN during the upsampling process. ResViT \cite{dalmaz2022resvit}, the first hybrid model designed for image-to-image translation, has shown excellent performance on several medical imaging tasks. Meanwhile, TransGAN \cite{jiang2021transgan}, which was limited by the inductive bias of ViT, developed a purely transformer-based generator. To address these challenges, TCGAN \cite{TCGAN} introduces a generator architecture that effectively combines a CNN and a transformer. This model successfully alleviates the small receptive field problem of CNN while preserving global feature transfer. EHNet \cite{EHNet} studies have shown that using convolution-based upsampling methods between transformer layers introduces unintended extraneous semantic information. This may damage the intended reconstruction. A series of studies have suggested that an effective CNN-transformer hybrid model can maintain semantic consistency.

\section{MRNET}
\subsection{Hierarchical SAM-Based Encoder}
\begin{figure*}[!t]
\centerline{\includegraphics[width=\textwidth, height=9cm]{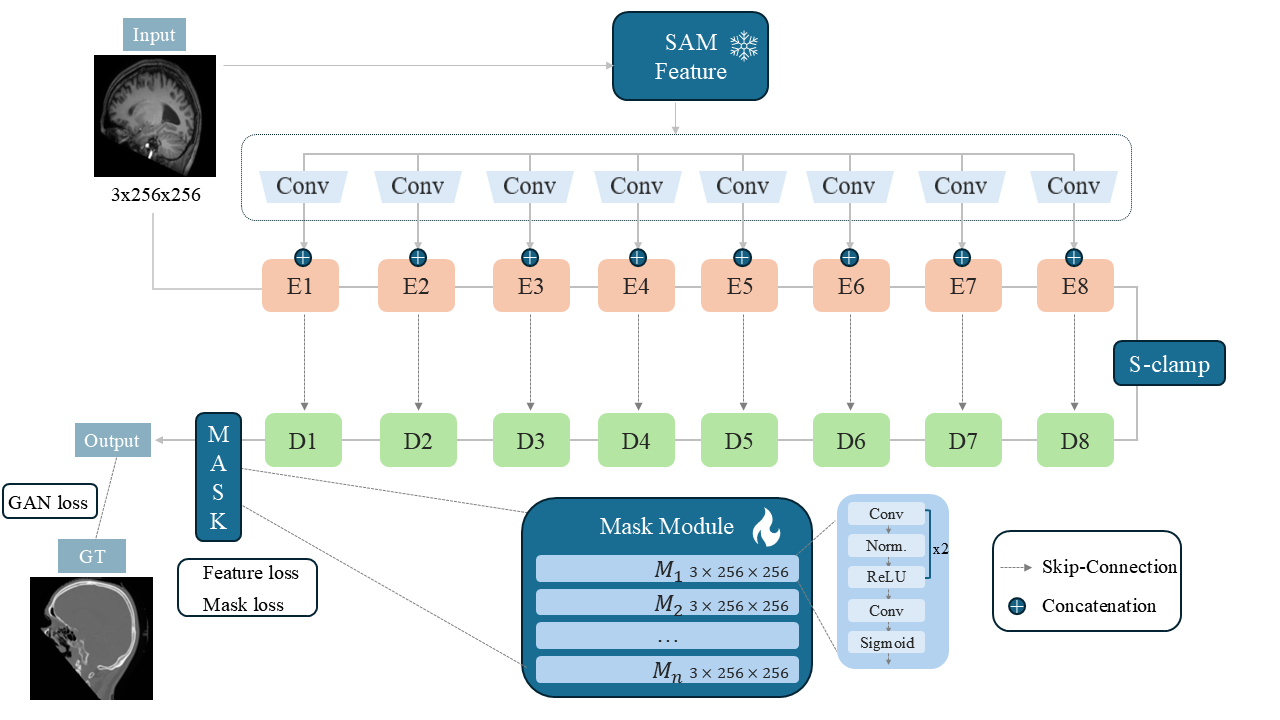}}
\caption{Illustration of generator G shows how the encoder network E processes a sampled feature map, which is then concatenated with the encoder features. The S-clamp facilitates a smooth transition of the feature at the lowest level between 0 and 1. Finally, the last stage of the decoder operates through the mask module, which is a multistage convolution-based component.}
\label{architecture}
\end{figure*}

Our generator is structured with a U-Net-like encoder-decoder architecture, which is depicted in Fig. \ref{architecture}.
Our study integrates pre-trained SAM image encoders to demonstrate their effectiveness in medical image representation and segmentation.
The MRNet encoder architecture merges SAM with a UNet-like structure and passes through a hierarchical encoder-decoder framework tailored for medical image transformation, focusing on capturing low-frequency information.
The feature integration process can be formally expressed as follows:

\begin{equation}
E_i = T_i(e_i \oplus A_i(featureSAM(x)))
\end{equation}

where $E_i$ represents the original encoder features, $A_i$ denotes the SAM adaptor function at stage $i$, and $\oplus$ indicates the channel wise concatenation. Transformation function $T_i$ encompasses stage-specific operations including normalization and nonlinear activation.

SAM adaptor $\mathcal{A}_i$ performs two crucial operations:

\begin{equation}
\mathcal{A}_i(featureSAM(x)) = \mathcal{C}_i(\mathcal{B}(F, s_i))
\end{equation}

Where $\mathcal{C}_i$ is the convolutional adaptation layer, $\mathcal{B}$ is bilinear interpolation, $\mathcal{F}$ is the SAM encoder feature and $s_i$ is the target spatial dimension at stage $i$. This formulation ensures both channel compatibility and spatial alignment between the SAM encoder features and the encoder representations.

The integration process starts by extracting features from the SAM model, which are passed through an adaptive convolution layer that acts as a SAM adapter. These adapted features are then bilinearly interpolated to fit the dimensions of the corresponding encoder stage. The interpolated features are then concatenated with the original encoder features along the channel dimensions. Finally, these combined features are then processed step-by-step to produce refined outputs for each encoder stage.

The integration mechanism is systematically applied to all encoder stages from E1 to E8, ensuring that the features are connected to the decoder branches at their respective resolutions via residual connections. This architectural design effectively captures the hierarchical features at various frequency levels while maintaining structural alignment with the input image. In particular, it prioritizes the extraction of incremental information at low frequencies, which enhances the model’s ability to transform detailed and accurate medical images while preserving essential structural information. By integrating SAM features, the model achieves improved performance in medical image transformation tasks while preserving important anatomical details and structural integrity. This design facilitates the integration of fine (local) and coarse (global) information even within deep structures. As a result, it reduces information loss and effectively resolves the tradeoff between details and semantic relationships.

\begin{figure}[!t]
\centerline{\includegraphics[width=\columnwidth]{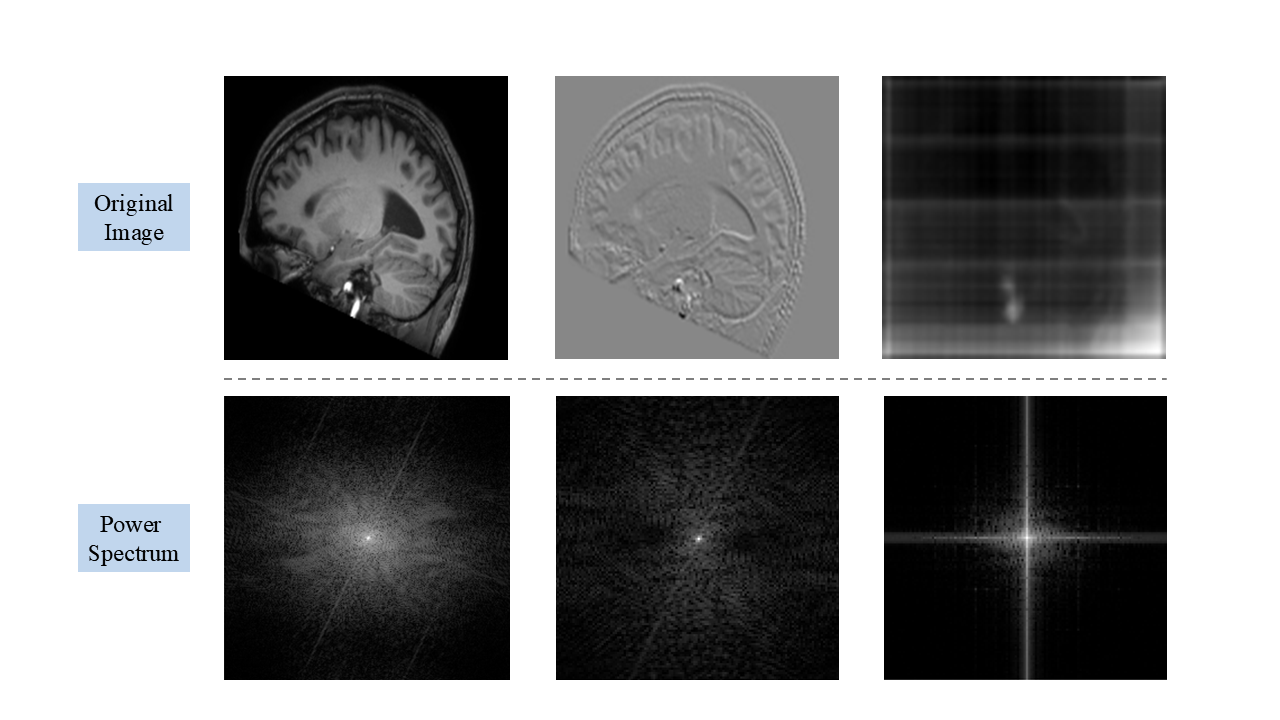}}
\caption{(Left) Original image (Middle); Image from the first layer of the encoder; (Right) Sum of the first layer of the encoder and the SAM feature received by the first layer of the decoder. 
}
\label{samlowfreq}
\end{figure}

The intermediate steps after feature extraction via SAM are visualized as shown in Fig. \ref{architecture}.
The power spectrum clearly exhibits the frequencies on which the model is trained. The bright spot in the center represents low-frequency components, whereas the outer part represents high-frequency components with detailed information. This suggests that the SAM is particularly sensitive to low-frequency components. The concentration of brightness in the central region and the darkness of the outer high-frequency area indicate a potential bias towards capturing broader, larger-scale features, while potentially neglecting detailed texture, which are then passed to the decoder.
As shown in the middle column of Fig. \ref{samlowfreq}, the output from the first encoder layer displays a balanced distribution of the high- and low-frequency components. When combined with the SAM and passed on to the decoder, the low-frequency components are filtered out and passed on, as depicted in the right column of Fig. \ref{samlowfreq}.

The final stage of our encoder pipeline incorporates a soft clamping mechanism\cite{monnier2021unsupervised, wu2024stegogan} to maintain data stability before decoder processing. This SoftClamp function is mathematically expressed as follows:

\begin{equation}
\begin{aligned} v &= \min(1, \max(0, z)) \\
\text{S}(z) &= v + \alpha(\min(0,z) + \max(0,z-1))\end{aligned}
\end{equation}

The SoftClamp function, denoted as $S(x)$, utilizes a scaling factor $\alpha$ with a softness parameter of 0.001 to clamp the feature values within the desired range, ensuring a smooth and stable transition to the decoder stage.

Our decoder architecture is intended to avoid complex attention mechanisms by implementing a lightweight design based on convolutional operations. This design offers several advantages in terms of resource efficiency and functional preservation. Given that the encoder already incorporates a computationally intensive SAM model, we intentionally balance the computational requirements of the network. Therefore, we use a lightweight decoder that relies on the underlying convolutional operations to maintain the overall performance of the network. This strategy is consistent with the work of \cite{unet}, which showed that complex decoder architectures do not necessarily lead to proportional improvements in performance. 

We achieve commendable performance through a robust hierarchical transformer encoder and a decoder composed solely of MLPs, especially when paired with a powerful encoder. Such as SegFormer\cite{segformer}. This approach highlights that a strong encoder can deliver accurate and high-resolution segmentation results without requiring complicated decoder designs. Thus, we focus on feature extraction using a vigorous hierarchical encoder that effectively captures both coarse and fine features across various resolutions while maintaining simplicity throughout the rest of the network.

The decoder architecture can be formally described as follows: At each decoder stage i (i $\in$ {1,...,8}), the output $D_i$ is computed through a series of operations involving skip connections from the encoder pathway. The feature maps from the encoder stages are concatenated with the upsampled decoder features along their channel dimensions. This concatenated output undergoes a transposed convolution followed by instance normalization and is subsequently processed through a ReLU activation function. The final decoder output is transformed through a hyperbolic tangent activation function before the mask correction stage.
This step captures the hierarchical feature processing and skip connection architecture characteristic of modern encoder-decoder networks, where information from the corresponding encoder levels is systematically integrated into the decoder pathway to preserve the fine-grained spatial details.

The discriminator and its associated loss function were implemented using the same version of the patch discriminator, as outlined in Pix2Pix\cite{pix2pix2017}. The patch size was set to 30, which was tailored to the dimensions of the training images employed in the present research.

\subsection{Dual Mask-Based Correction}

We adopted the mask generation mechanism from StegoGAN\cite{wu2024stegogan} and extended it into a multimask framework to enhance the selective feature extraction process during image translation. The generator employs three parallel masks, each generated through a dedicated convolutional neural network pathway that learns to identify and segment relevant features from the intermediate feature representations.

The mask generation process can be formally expressed as the following:

\begin{equation}
M_i = \sigma(\psi_i(\hat{y}_{i-1})), \quad i \in \{1, \ldots,n\}
\end{equation}

Where $M_i$ represents the $i$-th mask, $\psi_i$ denotes a sequence of convolutional layers with instance normalization and ReLU activation, and $\sigma$ is the sigmoid activation function that normalizes the mask values to the range of [0, 1]. When $i=1$, $y_{0}$ is the output of the generator's decoder pathway of the generator after tanh activation. 

These generated masks are then applied sequentially to modulate the output features:

\begin{equation}
\hat{y}_k = \hat{y}_{k-1} \odot \prod_{i=1}^k \sigma(\psi_i(\hat{y}_{k-1})), \quad k \in \{1,\ldots,n\}
\end{equation}
The process involves taking the translated values of $y_2$ and each mask $M_i$, which are sequentially applied. This method allows for iterative refinement and enhances the accuracy of the model in achieving the desired visual outcomes. More importantly, both the final translated output ($y_2$) and the mask values ($M_i$) are incorporated into the total loss computation, with the masks being compared against their respective ground truth masks for optimal feature selection.

This design enables each mask to automatically learn to identify and focus on different relevant features during the translation process. The effectiveness of the multimask mechanism was validated through quantitative comparisons with ground truth images, demonstrating improved preservation of anatomical structures and enhanced feature selection compared with single mask approaches. The masks adaptively adjust their attention patterns based on the input context, allowing for dynamic feature emphasis depending on the specific characteristics of each medical image.

\subsection{Overall Losses}

The total generator loss comprises multiple components designed to ensure high-quality image translation while maintaining structural consistency. Each component serves a specific purpose in the training process.

To ensure that the generator produces realistic outputs that can fool the discriminator, we employ an MSE-based adversarial loss:

\begin{equation}
\mathcal{L}_{GAN}(G,D) = \mathbb{E}_{x,y}[(D(G(x),x) - 1)^2]
\end{equation}

To maintain the structural consistency between the generated output and the ground truth, we utilize an L1 distance-based pixel-wise loss:

\begin{equation}
\mathcal{L}_{pixel}(G) = \mathbb{E}_{x,y}[\|y - G(x)\|_1]
\end{equation}

For each mask application step, we also compute the inverse masks and the corresponding discarded features:

\begin{equation}
\overline{M}_i = 1 - M_i
\end{equation}

The feature discarding loss penalizes activation in nonrelevant regions identified by the inverse masks:

\begin{equation}
\mathcal{L}_{feature} = \sum_{i=1}^{2}\|\hat{y}_i \odot \overline{M}_i\|_1
\end{equation}

\begin{equation}
\mathcal{L}_{mask} = \sum_{i=1}^{2}\|\overline{M}_i\|_1
\end{equation}
where $\hat{y}_i$ represents the features in each masked region. The feature discarding loss $L_{feature}$ penalizes activation in non-relevant regions identified by the inverse masks ($1$ - $M_i$), whereas the mask sparsity loss $L_{mask}$ promotes efficient attention distribution by encouraging masks to be selective in their activation to promote efficient attention distribution.
The final generator loss combines all components with the appropriate weighting factors:

\begin{equation}
\begin{aligned}
\mathcal{L} &= \mathcal{L}_{GAN}(G,D) + \lambda_{pixel}\mathcal{L}_{pixel}(G) \\
&\quad + \lambda_{feature}\mathcal{L}_{feature}(G) + \lambda_{mask}\mathcal{L}_{mask}(G)
\end{aligned}
\end{equation}

where $\lambda_{pixel} = 100$  controls the importance of the pixel wise reconstruction, $\lambda_{feature} = 0.1$ weights the feature preservation loss, and $\lambda_{mask} = 0.05$ balances the mask sparsity constraint.

This comprehensive loss function guides the network to achieve high-quality image translation while maintaining structural consistency and efficient feature selection through the multimask mechanism.

\section{Experiments}
\subsection{Datasets}
\subsubsection{MRI-to-CT Datasets}
The current study employed the MRI-CT paired dataset presented in the SynthRAD2023 challenge\cite{synthrad2023} for image-to-image translation. The dataset consists of 3D brain MRI-CT pairs, and we complied the dataset with the official protocols provided by SynthRAD2023 regarding image registration and preprocessing. The dataset, which included 180 subjects, was randomly divided into training, validation, and testing sets in an 8:1:1 ratio, yielding 144, 18, and 18 subjects for each set, respectively. Following the parsing of each slice along the x, y, and z axes, the dimensions of the 3D voxel-shaped images were resized to 256x256 pixels. In this learning approach, as opposed to the conventional method that primarily focuses on the z-axis slice, we gain the advantage of comprehensively examining the coronal, sagittal, and axial perspectives of the human body in a three-dimensional context. This methodology provides benefits such as data augmentation. Therefore, a total of 119,317 pairs of 2D image datasets were used in the study, the details of which are shown in Table \ref{dataset_distribution}.

\begin{table}[hbt!]
\renewcommand{\arraystretch}{1.3}
\caption{DATASET DISTRIBUTION OF MRI/CT PAIR}
\label{dataset_distribution}
\centering
\begin{tabular}{lcccc}
\hline
\bfseries & \bfseries Train & \bfseries Val & \bfseries Test & \bfseries Total \\
\hline
\hfill Paired CT/MRI & 95,559 & 11,987 & 11,771 & 119,317 \\
\hline
\end{tabular}
\end{table}

\subsubsection{MRI to MRI Datasets}
This study utilized the MRI-MRI paired dataset presented in Brats\cite{menze2014multimodalbrats1, bakas2017brats17, bakas2018ibrats17}. This dataset comprises images collected through clinical protocols and scanners from various institutions, including a total of 377 subjects, with multiple training and validation sets. Each patient had T1-weighted, T2-weighted, post-contrast T2-weighted, and T2 Fluid Attenuation Inversion Recovery (FLAIR) brain MR images. We conducted an experiment to convert the FLAIR modality to the T2 modality, resulting in a total of 371 person's axial cross-sections for training purposes. For testing, we employed 127 person's axial cross-sections. Ensuring that all variables were configured to match those used in the MRI-to-CT translation experiments.

\subsection{Experimental Settings}
In our experimental settings, the input images were resized to a tensor configuration of 3x256x256 and processed individually. To prevent overfitting, the dataset was meticulously divided into training, validation, and testing subsets following the established 8:1:1 ratio. The random seed was fixed in 2024 to ensure the reproducibility of the experiment. The Adam optimizer was initially configured with a learning rate of 0.0001 and momentum parameters of 0.5 and 0.999, which enhanced the gradient stability during training. The training duration encompassed a total of 40 epochs with a consistent batch size of 4. All experiments were conducted using the PyTorch framework, leveraging either four GPUs or a single A100 GPU. 

Competitive models, except for CycleGAN, were utilized as architecture generators within the c-GAN framework. All variables were held constant across MRNet and the discriminator, except for specific parameters tailored to each model. The models compared included Pix2Pix\cite{pix2pix2017}, CycleGAN\cite{CycleGAN2017}, TCGAN\cite{TCGAN}, TransUNet\cite{chen2021transunet}, SwinTransformer(Swin-T)\cite{liu2021swin}, ResViT\cite{dalmaz2022resvit} and PPT\cite{ppt}. All architectures, apart from CycleGAN, were trained using paired datasets, with loss updated through direct comparisons between ground truth and translated values. In similar studies, TransUNet and Swin-T replaced the segmentation head with a convolutional layer, following the methodology applied in \cite{dalmaz2022resvit}.
To ensure a fair comparison with competitive methods, all models utilized the officially distributed code. Based on the recommended configurations, the parameters were adjusted to optimize the performance according to the specific dataset. Early stopping techniques were also implemented to mitigate the overfitting.

\subsection{Result Analysis}
We evaluated the peak signal-to-noise ratio (PSNR) and structural similarity index (SSIM) for an objective and quantitative assessment. Using paired samples, we tested the translated B modal generated from the A modal learned during training, against the corresponding ground truth B modal image.

Table \ref{sotapair} proviedes a comprehensive comparison of various image enhancement models, based on PSNR and SSIM. These standard metrics, sidely used in image processing, indicate better image quality and structural fidelity with higher values.
\begin{table}[hbt!]
\renewcommand{\arraystretch}{1.3}
\caption{COMPARISON OF DIFFERENT MODELS (MRI -> CT) }
\label{sotapair}
\centering
\begin{tabular}{lccc}
\hline
\bfseries Model & \bfseries PSNR & \bfseries SSIM & \bfseries hybrid model \\
\hline
Pix2Pix & \underline{28.2379} & 0.7277 & \\
CycleGAN & 15.6958 & 0.4311 & \\
TCGAN & 23.3101 & 0.7688 & \checkmark \\
TransUNet & 23.5128 & 0.7558 & \checkmark \\
Swin-T & 24.3163 & 0.7717 & \ \\
ResVIT & 24.8740 & \underline{0.7896} & \checkmark \\
PPT & 7.5778 & 0.2351 &  \\
OURS & \textbf{29.9598} & \textbf{0.8977} & \checkmark \\
\hline
\end{tabular}
\end{table}

Among the existing methodologies, Pix2Pix performs well, with a PSNR of 28.2379 and an SSIM of 0.7277. It significantly outperforms the CycleGAN architecture, which has a PSNR of 15.6958 and an SSIM of 0.4311. This performance gap suggests that the direct image-to-image translation strategy used by Pix2Pix is more effective for the task compared to the cycle-consistent learning approach. Hybrid models that integrate the transformer architecture (TCGAN, TransUNet, and ResVIT) outperform CycleGAN, with a PSNR value of 23.3101–24.8740 and a SSIM value of 0.7558–0.7896. These trends generally suggest that incorporating the transformer component improves the model’s ability to capture long-range dependencies and complex image features. A pure Transformer-based model, Swin-T, shows an intermediate result, with a PSNR of 24.3163 and an SSIM of 0.7717. This suggests that while the transformer architecture can adeptly manage image enhancement tasks, a hybrid strategy may provide more benefits in balancing local and global feature processing.

Most importantly, the proposed model outperforms all the baseline models by achieving a PSNR of 30.7189 and an SSIM of 0.9111. Compared to the second best performing ResVIT, our approach achieves a PSNR of 5.8449 and SSIM of 0.1215. This significant improvement is attributed to the innovative design of our model architecture. The consistently superior performance in both metrics strongly validates our architectural design choices and demonstrates the effectiveness of our approach in preserving both pixel-level accuracy (denoted by PSNR) and structural consistency (denoted by SSIM). These quantitative results confirm that the proposed model achieves state-of-the-art performance, especially in scenarios where detailed preservation and structural integrity are most important.

\begin{figure*}[!t]
\centerline{\includegraphics[width=\textwidth]{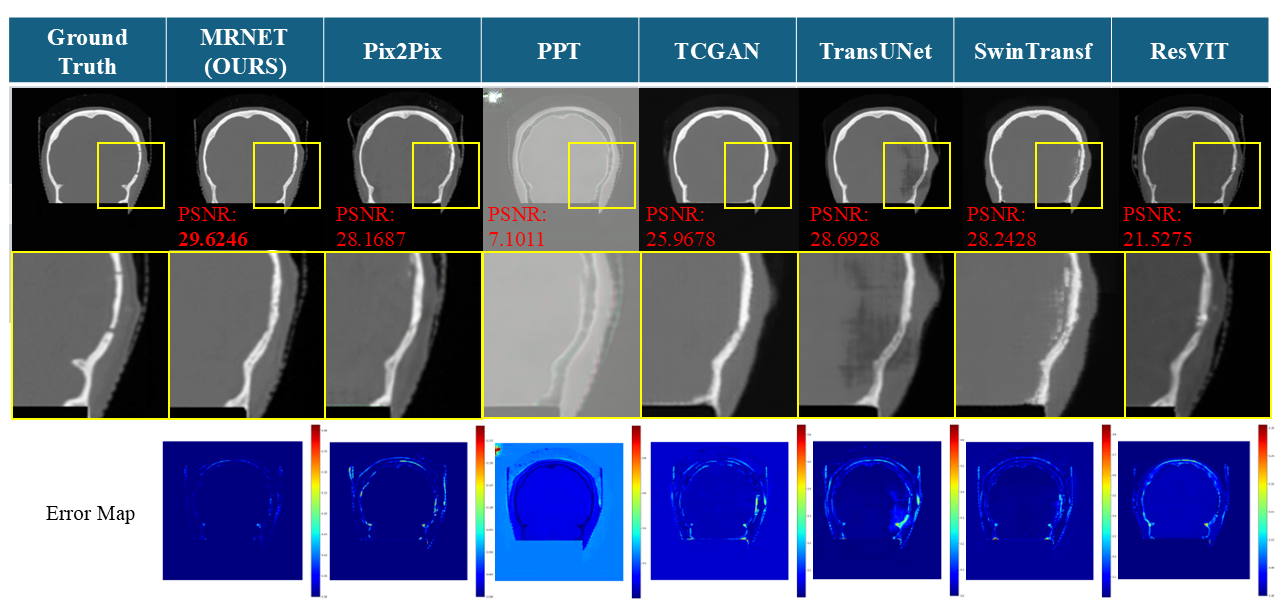}}
\caption{This visualization presents the results for the z-axis of the image, which has been translated into the CT modality using the MRI input. MRNet displayed the highest performance based on individual PSNR values. Unlike other methods, which resulted in a jittery appearance or inaccurately rendered bone structures in areas where they should not appear, our approach demonstrated a clean and stable translation. The top row displays the translation results, whereas the middle row offers an enlarged view of the area highlighted in the yellow box. The error map at the bottom clearly indicates that our method, which exhibited the least activation, produced the most stable outcome.}
\label{viszaxis}
\end{figure*}

\begin{figure*}[hbt!]
\centerline{\includegraphics[width=\textwidth]{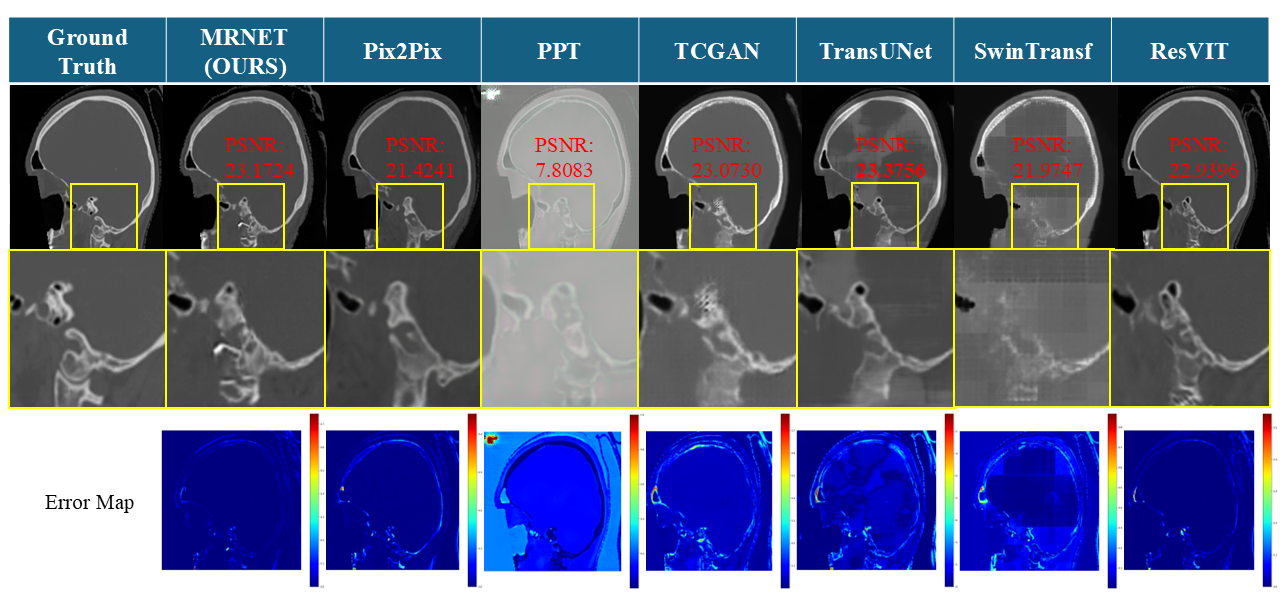}}
\caption{This visualization presents the results for the y-axis of the image, which has been translated into the CT modality using the MRI input. The top row depicts the translation results. The middle row provides an enlarged view of the area highlighted within the yellow box. The differences from the ground truth are illustrated in the error map shown at the bottom. MRNet achieved the second highest performance in terms of individual visualization of the PSNR values. However, the visualizations show that it is closer to the ground truth than the first-performing method. The transformer-only method exhibited grid-like and mottled features.
}
\label{visyaxis}
\end{figure*}

Fig. \ref{viszaxis} and Fig. \ref{visyaxis} show the visualization results of the MRI-CT transformation using the ground truth of the target domain. We sliced the images along the x, y, and z axes to differentiate them from the general learning process and enhance the effectiveness of data augmentation. Similarly, we performed tests in all dimensions and show the selected visualizations.
The experimental results show the performance differences by anatomical axis in cross-modal medical image transformation. In the z-axis analysis shown in Fig. \ref{viszaxis}, the proposed method achieves a PSNR of 29.6246, significantly outperforming the second-ranked TransUNet by a difference of 28.6928. This outstanding performance is particularly evident in the transformation of MRI images to the CT modality.

The qualitative analysis of the z-axis transformation shows that competing methods often introduce artifacts, resulting in irregular intensity fluctuations and anatomically incorrect bone structure representations. In contrast, our method consistently produces stable and anatomically accurate transformations. The comparative visualization protocol, consisting of the transformation results (top row) and the enlarged region of interest (middle row and yellow box), provides clear evidence of this superiority. Furthermore, the error distribution maps (bottom row) quantitatively support these observations. Our method shows minimal activation patterns, indicating excellent stability in the transformation process.

For the y-axis (sagittal) transformation, the individual quantitative metrics in Fig. \ref{visyaxis} show that our method achieves a PSNR of 23.1724, which is slightly lower than the leading TransUNet performance of 23.3756. However, a detailed visual analysis revealed important qualitative differences. During the MRI-to-CT transformation process, the y-axis transformation showed a noticeable structural degradation in the cervical vertebrae representation, although TransUNet achieved a nominally better PSNR score.

The visualization protocol for the y-axis analysis is consistent with the z-axis presentation format, featuring full-field transformation, ROI augmentation, and error distribution maps. The transformation of the MRNet shows excellent fidelity to real anatomical structures compared to the top-ranked methods. A notable observation in the visualization is that the transformer-based method consistently produces characteristic lattice artifacts and heterogeneous texture patterns that can be attributed to the structural characteristics. These findings suggest that it is important to perform both quantitative metrics and qualitative analyses for cross-modal medical image conversion performance to have clinical significance.

We focused on the MRI-MRI dataset and evaluated the efficacy of MRNet in the process of converting FLAIR images to T2 images. To ensure the reliability of the results, we included the same model used in the MRI-CT conversion task before conducting the second experiment. In this experiment, we used both qualitative and quantitative metrics. As presented in the Table \ref{bratspair}, the effectiveness of the medical image translation technology used was more accurately verified.

\begin{table}[hbt!]
\renewcommand{\arraystretch}{1.3}
\caption{COMPARISON OF DIFFERENT MODELS (FLAIR -> T2)}
\label{bratspair}
\centering
\begin{tabular}{lccc}
\hline
\bfseries Model & \bfseries PSNR & \bfseries SSIM & \bfseries Hybrid model \\
\hline
Pix2Pix & 25.3255 & 0.8957 & \\
CycleGAN & 25.0428 & 0.8906 & \\
TCGAN & 29.5175 & 0.9388 & \checkmark \\
TransUNet & 29.5681 & \underline{0.9396} & \checkmark \\
Swin-T & \underline{30.2356} & 0.9337 & \ \\
ResVIT & 26.7941 & 0.9174 & \checkmark \\
PPT  & 6.2076 & 0.0711&  \\
OURS & \textbf{31.2852} & \textbf{0.9411} & \checkmark \\
\hline
\end{tabular}
\end{table}

Among the existing methodologies, Pix2Pix and CycleGAN show similar performance. Pix2Pix achieves a PSNR of 25.3255 and an SSIM of 0.8957, whereas CycleGAN achieves a PSNR of 25.0428 and an SSIM of 0.8906. The relatively small performance difference between the two models suggests that direct image-to-image translation and cycle-consistency learning approaches are similarly effective for the task of converting FLAIR to T2. Hybrid models that incorporate transformer architectures, such as TCGAN, TransUNet, and ResVIT, show various performances. TCGAN and TransUNet show notable improvements, achieving PSNR values of 29.5175 and 29.5681 and SSIM values of 0.9388 and 0.9396, respectively. This highlights that the hybrid models are effective in preserving soft tissue structural information of MRI during the conversion process. The pure transformer-based model Swin Transformer outperforms most hybrid approaches with PSNR of 30.2356 and SSIM of 0.9337. This suggests that the transformer architecture is particularly suitable for capturing the complex relationship between the FLAIR and T2 modalities. Interestingly, the performance of ResVIT is moderate with a PSNR of 26.7941 and an SSIM of 0.9174.

Most importantly, the proposed model outperforms all the compared models with PSNR of 31.2852 and SSIM of 0.9411. Although these improvements are smaller compared to the MRI-to-CT conversion task, they still demonstrate the robustness and effectiveness of our architectural design.
The patch-based PPT model showed significantly lower performance indices and was not suitable for this particular conversion task. These quantitative results comprehensively validate the effectiveness of our model in preserving both pixel-level accuracy and structural consistency in the FLAIR-to-T2 conversion task, establishing a new state-of-the-art benchmark in this domain.

\subsection{Ablation Study}
We conducted different ablation studies to verify the effects of the multimask frameworks and specialized loss functions on the MRI-to-CT dataset. The first ablation study was conducted by varying the number of masks. Then, we tested the effect with and without mask loss.

\subsubsection{Amount of Mask Ablation}
The proposed algorithm utilizes multiple masks to minimize the disparity between the MR-based generated CT image and the ground truth CT. We conducted an ablation study to confirm the effectiveness of the mask algorithm. Our findings confirm that employing multiple masks leads to better performance than using no masks or just one mask. Table \ref{ablation_mask} contains information about the number of masks employed to achieve the most significant performance enhancement.

\begin{table}[hbt!]
\renewcommand{\arraystretch}{1.3}
\caption{TEST PERFORMANCE DEPENDING ON THE NUMBER OF MASKS}
\label{ablation_mask}
\centering
\begin{tabular}{lccc}
\hline
\bfseries Number of masks (M$_n$) & \bfseries PSNR & \bfseries SSIM \\

\hline
\hfill 1 (M$_1$) & 28.3605 & 0.8845 & \\
\hfill 2 (M$_2$) & \textbf{29.9598} & \textbf{0.8977} & \\
\hfill 3 (M$_3$) & 29.6911 & 0.8893 & \\
\hfill 4 (M$_4$) & 20.7559 & 0.8094 & \\
\hfill 6 (M$_6$) & 20.4624 & 0.7585 & \\
\hfill 7 (M$_7$) & 18.2125 & 0.6373 & \\

\hline
\end{tabular}
\end{table}

The use of multimasks was determined through extensive experimentation, demonstrating that this configuration provides an optimal balance between model complexity and translation quality. One mask resulted in insufficient feature capture, whereas additional masks led to redundant attention patterns without meaningful performance gains.

\subsubsection{Incorporation of Mask-Based Loss}
The integration of the dual-mask-based loss into our training framework has enhanced both the stability and performance of the model. Mask loss, derived from segmentation or object masking processes, was introduced to refine the image quality by directing the focus of the model toward critical regions. The goal was to ensure that the model not only captures global features through adversarial loss but also incorporates local contextual and structural information through mask guidance.

\begin{table}[hbt!]
\renewcommand{\arraystretch}{1.3}
\caption{TEST PERFORMANCE DEPENDING ON FEATURE LOSS AND MASK LOSS}
\label{result_table}
\centering
\begin{tabular}{lccc}
\hline
\bfseries Model & \bfseries PSNR & \bfseries SSIM &  \\
\hline
A. Baseline(GAN loss) & 28.4141 & 0.8495 & \\
B. A + feature loss + mask loss  & \textbf{29.9598} & \textbf{0.8977}  \\
\hline
\end{tabular}
\end{table}

Table \ref{result_table} quantitatively demonstrates the effect of introducing mask loss. It compares the performance of the models trained with and without this loss. The baseline model (Model A) using only GAN loss achieves a PSNR of 28.4141 and an SSIM of 0.8495. These results demonstrate the basic capabilities of the network when there is no mask loss. Model B, which integrates both GAN loss and feature and mask loss, shows significant improvement, achieving a PSNR of 29.9598 and an SSIM of 0.8977.

These improvements are a result of the effective guidance of the mask-based loss function of the model focusing on important regions within the image and minimizes errors in those regions. Through mask-based feedback, the model learns to reduce global errors and local mismatches that may be overlooked in standard adversarial training. As a result, it reconstructs the image structure and texture more accurately and improves fidelity to the original content.

These findings suggest that the mask loss is essential for the model to effectively balance global and local feature representations.
It is especially useful for tasks that require high structural accuracy and semantic consistency.

\section{CONCLUSION AND DISCUSSION}

The current research introduces MRNet, an innovative architecture for medical image-to-image translation, addressing longstanding challenges in the field. By integrating the frequency analysis capabilities of SAM with a dual-mask framework, MRNet achieves state-of-the-art performance in MRI-to-CT and MRI-to-MRI translation.
The model was trained using a paired dataset that provides a robust baseline performance. However, given the limited availability of well-structured paired datasets in the medical domain, future research should focus on developing models capable of achieving strong performance with minimal paired datasets.

A key strength of MRNet is that it is based on a well-established UNet-like encoder-decoder architecture, enabling easy module-specific modifications for enhanced performance. This study aims to inspire the development of lightweight models capable of delivering accurate results. While MRNet performs reliably, opportunities remain for optimization, particularly in computational efficiency and edge case handling in clinical settings. The success of the network opens up several promising research avenues, especially within a clinical context. For instance, plane-aware translation mechanisms could address performance variations across specific planes. Additionally, incorporating clinical validation metrics should be incorporated to ensure alignment with real-world medical applications.

\bibliographystyle{IEEEtran}
\bibliography{references}

\end{document}